\documentclass[sigconf]{acmart}

\AtBeginDocument{%
  }

\setcopyright{None}

\acmConference[FAccTRec '24]{7th FAccTRec Workshop on Responsible Recommendation}{October 14, 2024}{Bari, Italy}

\begin{document}

\title{RTs $\neq$ Endorsements: Rethinking Exposure Fairness on Social Media Platforms}

\author{Nathan Bartley}

\email{nbartley@usc.edu}
\author{Kristina Lerman}
\affiliation{%
  \institution{Information Sciences Institute}
  \city{Marina Del Rey}
  \state{California}
  \country{USA}
}

\renewcommand{\shortauthors}{Bartley \& Lerman}

\begin{abstract}
  Recommender systems underpin many of the personalized services in the online information \& social media ecosystem. However, the assumptions in the research on content recommendations in domains like search, video, and music are often applied wholesale to domains that require a better understanding of why and how users interact with the systems. In this position paper we focus on social media and argue that personalized timelines have an added layer of complexity that is derived from the social nature of the platform itself. In particular, definitions of exposure fairness should be expanded to consider the social environment each user is situated in: how often a user is exposed to others is as important as \textit{who} they get exposed to. 
\end{abstract}




\maketitle

\section{Introduction}

Solving for fairness in recommender systems is a difficult problem with multiple types of stakeholders, e.g., consumers, producers, and advertisers \cite{ekstrand2022fairness}. As with other algorithmic systems, there has been acknowledgement of harms from recommender systems, both allocative and representative \cite{barocas2017problem}. Many concerns around algorithmic harms stem from the fact that we measure performance of recommender systems by aggregating over the score of each user's individual recommendations rather than a top-down view of the behavior of the system. These measures of bias and harm are useful for informing better recommender system design, but a common theme in these analyses is a tacit abstraction of the domain that the system is situated in \cite{ekstrand2022matching,ekstrand2022fairness}. 

Take for instance popularity bias: the observation that popularity bias in music recommendations can yield material differences in revenue across artists is useful for mitigating those harms in the music recommendation domain and only marginally useful for other domains of recommendations. Popularity bias in social media personalized timelines can also have material consequences for content creators in terms of revenue, however the social context of the recommendations means that a user's engagement with content is also dependent on other factors. These factors include one's own beliefs or ideological alignment to the content; how one perceives the norms about engaging with the content; one's perceived alignment with the user sharing the content; finally, any considerations the individual has for their ``imagined audiences'' and how they might be perceived for engaging with the content\cite{marwick2011tweet}. These factors may be present in other kinds of recommendation domains, however the extent to which they are present depends on the structure and affordances of the platform. 

Another aspect of this is how to enforce constraints about allocative harms like exposure fairness in recommender systems. Extensive work has been done to address these concerns, however they tacitly assume the inventory of items is universally accessible to each user, and also do not consider systems that rely on user-generated content (i.e., where consumers are also producers)\cite{wu2022joint,diaz2020evaluating,burke2018balanced,abdollahpouri2020multi}. Given the network structure underpinning social media platforms, who one user is connected to will have substantial influence on the content they are exposed to. In order to enforce constraints about exposure in social media recommender systems, it is important for such measures of fairness  to be designed with users' perceptions, possible cognitive biases and their interactions with other users (not to mention the system itself) in mind.

Given the above caveats, in this position paper we build upon previous arguments that metrics and fairness strategies should be adapted to the domain they are situated in \cite{ekstrand2022matching}. We discuss the social media personalized timeline context specifically, the social and cognitive factors impacting perception, and what exposure bias entails. We then present a toy example of exposure bias and suggest a means for mitigation.

\section{Social sensing \& cognitive biases}

First we consider the users' perceptions and biases. Human beings are fundamentally social creatures, and it has been shown that we have the capacity to observe our social networks and make inferences about the mental states of others. This capacity for social sensing is susceptible however to cognitive biases: Galesic, Olsson and Rieskamp 2012 describe a model where individuals estimate large scale statistics like average household wealth by using information from their immediate social network \cite{galesic2012social}. In this they found individuals surveyed tended to be more accurate when making estimates about their immediate network, but were much less accurate when making estimates about broader populations.

We refer to the cognitive sciences to identify relevant socio-cognitive biases. A primary bias is the salience bias: humans tend to pay more attention to unexpected or irregular stimuli \cite*{kardosh2022minority}. This bias can manifest in identifying minority groups as standing out, often resulting in an overestimation of their size. This is similar to the structural phenomena in networks in which network structure distorts a user's local perception of the whole network, e.g., the majority illusion and the friendship paradox\cite{lerman2016majority}. 

The individual's own prior beliefs can influence their social perception as well through the false uniqueness (i.e., underestimating the prevalence of one's views) and false consensus effects (i.e., overestimating the prevalence of one's views). Empirical evidence suggests that Americans with more conservative views on climate change, as well as those with more conservative local norms and exposure to conservative news underestimate support for climate change policies by as much as half \cite{sparkman2022americans}. These biases have also been shown to be important to overcoming collective action problems in theoretical games \cite{santos2021biased}. 

\section{Operationalization of Exposure Bias}

As a minority/majority dichotomy in a social network can reflect a wide range of factors from demographic factors like gender, ethnicity and age to differences in opinion on arbitrary matters, we consider \textit{exposure bias} to be a systematic distortion in content visibility and perceived prevalence within a social network. This distortion arises from discrepancies between the ``potential network'' (active social connections, e.g., who follows who), the ``activated network'' (the users who engage in content creation or activity and thus can be observed by other users), and the feed-exposed network (the network users actually observe and interact with). The feed-exposed networks, mediated by specific recommender systems, can modulate this bias leading to certain content being either over- or under-represented, which skews its perceived prevalence relative to its actual prevalence. In a sense, minimizing this bias can be seen as enforcing statistical parity in exposure across users in the ``global'' network\cite{mehrabi2021survey}. 

There are multiple ways one may measure exposure bias– previous studies have used the Gini coefficient and local perception bias as introduced in Alipourfard et al., 2020 \cite{bartley2023evaluating,alipourfard2020friendship}. These measures can be adjusted to take exposure into consideration, allowing us to compare exposure under different personalized timelines. As described in previous works, it is essential to consider multiple metrics as they complement each other in interpreting their results\cite{mansoury2021graph}.

An important note in enforcing any measure of exposure fairness is that it may work well in aggregate, however depending on what ``global'' represents there may be subsets of users with vastly different experiences of their network. To better understand the heterogeneous experiences users may have of their network, we should consider how users interact with each other and with the platform itself.

\section{Interpersonal dynamics \& utility}

Three interpersonal user dynamics have been described in these online ``networked publics'' that should be considered when building personalized timelines: 1) unless explicitly constrained to friends and followers, users have invisible audiences to their content that may scale in size beyond their control; 2) the different contexts in which content is observed are collapsed  (e.g., the designated audience may perceive a joke to be appropriate but others may perceive the same content to be inappropriate); and 3) similar to the first point, content and conversations assuming more private environments may be thrust into more public environments with unintended consequences \cite{boyd2009social}. 

These considerations suggest that \textit{who} shows up on your timeline and the social context in which their post is presented can shape users' experiences of the platform. As a practical example, Instagram researchers have suggested that they filter out some rare user IDs from users' feeds as they have found too many rare users can impact users' satisfaction / utility with the system \cite{bredillet}. The measurements of user utility and relevance, tied into an optimization that rewards engagement, may inadvertently optimize a user's feed for niche tastes and specific behaviors. For instance, Jiang et al., 2023 suggests that toxic behavior on platforms like X/Twitter may be reinforced by social approval and the propensity to act in a toxic behavior in their network \cite{jiang2023social}. If we are not careful about user dynamics when enforcing fairness we may end up optimizing negative engagement between users or facilitating second-order effects like fewer professional connections for marginalized users\cite{akpinar2024authenticity}.

\section{Toy example \& Practical Considerations}

\begin{figure}
    \centering
    \includegraphics[width=\linewidth]{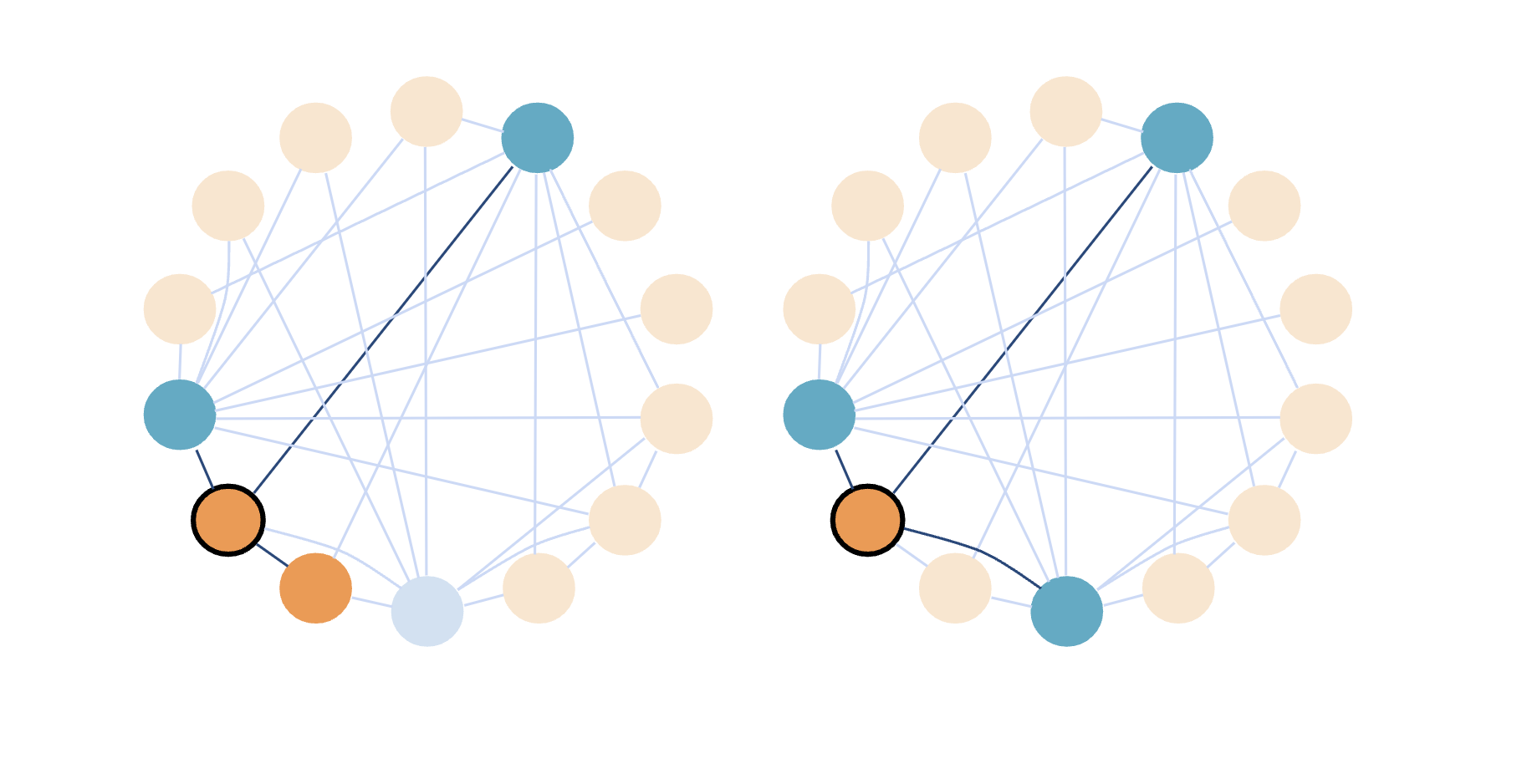}
    \caption{\textbf{Simple exposure example}. On the left the user is exposed to two users with a globally minority trait and one with the majority trait. On the right the same user, under a different feed is exposed to three users with the minority trait.}
    \label{fig:example}
\end{figure}

Consider the network described in Fig. \ref{fig:example}. In this example users are subjected to two different kinds of timelines with limited slots (i.e., each user only observes their timeline for a fixed amount of posts). One can imagine that in aggregate both feeds show each users' posts in a manner that maintains a particular measure of exposure fairness (e.g., the probability of observing a post from a user with the minority trait is proportional to their prevalence in expectation). The problem arises with the network: each user within that network will observe a different prevalence of the trait, and as such may have a different perception of whether or not the system is actually providing fairness in exposure. If the user highlighted in the figure is only exposed to blue users, then this may have consequences on how the user perceives the network as a whole.

To mitigate this exposure concern, practitioners in this may consider the homophilic structure of each user's network when selecting candidates for recommendation\cite{lee2019homophily}: the degree-attribute correlation $\rho_{kx}$ is well-understood to be connected to the strength of these phenomena like the majority illusion\cite{lerman2016majority}. This correlation is defined as follows:

\begin{align*}
    \rho_{kx} = \frac{P(x=1)}{\sigma_{x}\sigma_{k}}[\langle k \rangle_{x=1} - \langle k \rangle]
\end{align*}
where x is the binary attribute, k is the degree of the node (here in-degree), $\sigma_{x}, \sigma_{k}$ the standard deviations of the binary attribute and in-degree respectively, and $\langle k \rangle$ the average in-degree over all nodes. 

Keeping the effective $\rho_{kx}$ close to zero can be a feasible vector for minimizing the distorted perception across users. This may be readily inserted alongside more complicated recommendations based on user history and social dynamics described previously.

\section{Conclusion}

In this paper we argue that to make adequate and exposure-fair recommendations in social media the platforms need to consider the perceived social environment the system is putting the user into. We suggest an operationalization of exposure bias that considers how a recommender system can modulate a user's perception of their social environment. We also suggest a potential method for minimizing such exposure bias. 

Of course, to some degree the user has agency in what and who they choose to interact with, however the platform decides how users can interact with it through design choices and affordances of the platform suggesting they have more liability in how they shape users' experiences. In order to adequately understand exposure fairness on these platforms, we must connect it more to each users' view of their online surroundings. 

\bibliographystyle{ACM-Reference-Format}
\bibliography{manuscript}


\begin{thebibliography}{22}


\ifx \showCODEN    \undefined \def \showCODEN     #1{\unskip}     \fi
\ifx \showDOI      \undefined \def \showDOI       #1{#1}\fi
\ifx \showISBNx    \undefined \def \showISBNx     #1{\unskip}     \fi
\ifx \showISBNxiii \undefined \def \showISBNxiii  #1{\unskip}     \fi
\ifx \showISSN     \undefined \def \showISSN      #1{\unskip}     \fi
\ifx \showLCCN     \undefined \def \showLCCN      #1{\unskip}     \fi
\ifx \shownote     \undefined \def \shownote      #1{#1}          \fi
\ifx \showarticletitle \undefined \def \showarticletitle #1{#1}   \fi
\ifx \showURL      \undefined \def \showURL       {\relax}        \fi
\providecommand\bibfield[2]{#2}
\providecommand\bibinfo[2]{#2}
\providecommand\natexlab[1]{#1}
\providecommand\showeprint[2][]{arXiv:#2}

\bibitem[Abdollahpouri and Mansoury(2020)]%
        {abdollahpouri2020multi}
\bibfield{author}{\bibinfo{person}{Himan Abdollahpouri} {and} \bibinfo{person}{Masoud Mansoury}.} \bibinfo{year}{2020}\natexlab{}.
\newblock \showarticletitle{Multi-sided exposure bias in recommendation}.
\newblock \bibinfo{journal}{\emph{arXiv preprint arXiv:2006.15772}} (\bibinfo{year}{2020}).
\newblock


\bibitem[Akpinar and Fazelpour(2024)]%
        {akpinar2024authenticity}
\bibfield{author}{\bibinfo{person}{Nil-Jana Akpinar} {and} \bibinfo{person}{Sina Fazelpour}.} \bibinfo{year}{2024}\natexlab{}.
\newblock \showarticletitle{Authenticity and exclusion: social media recommendation algorithms and the dynamics of belonging in professional networks}.
\newblock \bibinfo{journal}{\emph{arXiv preprint arXiv:2407.08552}} (\bibinfo{year}{2024}).
\newblock


\bibitem[Alipourfard et~al\mbox{.}(2020)]%
        {alipourfard2020friendship}
\bibfield{author}{\bibinfo{person}{Nazanin Alipourfard}, \bibinfo{person}{Buddhika Nettasinghe}, \bibinfo{person}{Andr{\'e}s Abeliuk}, \bibinfo{person}{Vikram Krishnamurthy}, {and} \bibinfo{person}{Kristina Lerman}.} \bibinfo{year}{2020}\natexlab{}.
\newblock \showarticletitle{Friendship paradox biases perceptions in directed networks}.
\newblock \bibinfo{journal}{\emph{Nature communications}} \bibinfo{volume}{11}, \bibinfo{number}{1} (\bibinfo{year}{2020}), \bibinfo{pages}{707}.
\newblock


\bibitem[Barocas et~al\mbox{.}(2017)]%
        {barocas2017problem}
\bibfield{author}{\bibinfo{person}{Solon Barocas}, \bibinfo{person}{Kate Crawford}, \bibinfo{person}{Aaron Shapiro}, {and} \bibinfo{person}{Hanna Wallach}.} \bibinfo{year}{2017}\natexlab{}.
\newblock \showarticletitle{The problem with bias: Allocative versus representational harms in machine learning}. In \bibinfo{booktitle}{\emph{9th Annual conference of the special interest group for computing, information and society}}. New York, NY, \bibinfo{pages}{1}.
\newblock


\bibitem[Bartley et~al\mbox{.}(2023)]%
        {bartley2023evaluating}
\bibfield{author}{\bibinfo{person}{Nathan Bartley}, \bibinfo{person}{Keith Burghardt}, {and} \bibinfo{person}{Kristina Lerman}.} \bibinfo{year}{2023}\natexlab{}.
\newblock \showarticletitle{Evaluating Content Exposure Bias in Social Networks}. In \bibinfo{booktitle}{\emph{Proceedings of the International Conference on Advances in Social Networks Analysis and Mining}}. \bibinfo{pages}{379--383}.
\newblock


\bibitem[Boyd(2009)]%
        {boyd2009social}
\bibfield{author}{\bibinfo{person}{Danah Boyd}.} \bibinfo{year}{2009}\natexlab{}.
\newblock \showarticletitle{Social media is here to stay... now what}.
\newblock \bibinfo{journal}{\emph{Microsoft Research Tech Fest}}  \bibinfo{volume}{5} (\bibinfo{year}{2009}).
\newblock


\bibitem[Bredillet(2023)]%
        {bredillet}
\bibfield{author}{\bibinfo{person}{Thomas Bredillet}.} \bibinfo{year}{2023}\natexlab{}.
\newblock \showarticletitle{Large Scale Recommendations at Instagram}.
\newblock   \bibinfo{volume}{VideoRecSys 2023} (\bibinfo{year}{2023}).
\newblock
\urldef\tempurl%
\url{https://videorecsys.com/slides/thomas_talk1.pdf}
\showURL{%
\tempurl}


\bibitem[Burke et~al\mbox{.}(2018)]%
        {burke2018balanced}
\bibfield{author}{\bibinfo{person}{Robin Burke}, \bibinfo{person}{Nasim Sonboli}, {and} \bibinfo{person}{Aldo Ordonez-Gauger}.} \bibinfo{year}{2018}\natexlab{}.
\newblock \showarticletitle{Balanced neighborhoods for multi-sided fairness in recommendation}. In \bibinfo{booktitle}{\emph{Conference on fairness, accountability and transparency}}. PMLR, \bibinfo{pages}{202--214}.
\newblock


\bibitem[Diaz et~al\mbox{.}(2020)]%
        {diaz2020evaluating}
\bibfield{author}{\bibinfo{person}{Fernando Diaz}, \bibinfo{person}{Bhaskar Mitra}, \bibinfo{person}{Michael~D Ekstrand}, \bibinfo{person}{Asia~J Biega}, {and} \bibinfo{person}{Ben Carterette}.} \bibinfo{year}{2020}\natexlab{}.
\newblock \showarticletitle{Evaluating stochastic rankings with expected exposure}. In \bibinfo{booktitle}{\emph{Proceedings of the 29th ACM international conference on information \& knowledge management}}. \bibinfo{pages}{275--284}.
\newblock


\bibitem[Ekstrand et~al\mbox{.}(2022)]%
        {ekstrand2022fairness}
\bibfield{author}{\bibinfo{person}{Michael~D Ekstrand}, \bibinfo{person}{Anubrata Das}, \bibinfo{person}{Robin Burke}, \bibinfo{person}{Fernando Diaz}, {et~al\mbox{.}}} \bibinfo{year}{2022}\natexlab{}.
\newblock \showarticletitle{Fairness in information access systems}.
\newblock \bibinfo{journal}{\emph{Foundations and Trends{\textregistered} in Information Retrieval}} \bibinfo{volume}{16}, \bibinfo{number}{1-2} (\bibinfo{year}{2022}), \bibinfo{pages}{1--177}.
\newblock


\bibitem[Ekstrand and Pera(2022)]%
        {ekstrand2022matching}
\bibfield{author}{\bibinfo{person}{Michael~D Ekstrand} {and} \bibinfo{person}{Maria~Soledad Pera}.} \bibinfo{year}{2022}\natexlab{}.
\newblock \showarticletitle{Matching Consumer Fairness Objectives \& Strategies for RecSys}.
\newblock \bibinfo{journal}{\emph{arXiv preprint arXiv:2209.02662}} (\bibinfo{year}{2022}).
\newblock


\bibitem[Galesic et~al\mbox{.}(2012)]%
        {galesic2012social}
\bibfield{author}{\bibinfo{person}{Mirta Galesic}, \bibinfo{person}{Henrik Olsson}, {and} \bibinfo{person}{J{\"o}rg Rieskamp}.} \bibinfo{year}{2012}\natexlab{}.
\newblock \showarticletitle{Social sampling explains apparent biases in judgments of social environments}.
\newblock \bibinfo{journal}{\emph{Psychological Science}} \bibinfo{volume}{23}, \bibinfo{number}{12} (\bibinfo{year}{2012}), \bibinfo{pages}{1515--1523}.
\newblock


\bibitem[Jiang et~al\mbox{.}(2023)]%
        {jiang2023social}
\bibfield{author}{\bibinfo{person}{Julie Jiang}, \bibinfo{person}{Luca Luceri}, \bibinfo{person}{Joseph~B Walther}, {and} \bibinfo{person}{Emilio Ferrara}.} \bibinfo{year}{2023}\natexlab{}.
\newblock \showarticletitle{Social approval and network homophily as motivators of online toxicity}.
\newblock \bibinfo{journal}{\emph{arXiv preprint arXiv:2310.07779}} (\bibinfo{year}{2023}).
\newblock


\bibitem[Kardosh et~al\mbox{.}(2022)]%
        {kardosh2022minority}
\bibfield{author}{\bibinfo{person}{Rasha Kardosh}, \bibinfo{person}{Asael~Y Sklar}, \bibinfo{person}{Alon Goldstein}, \bibinfo{person}{Yoni Pertzov}, {and} \bibinfo{person}{Ran~R Hassin}.} \bibinfo{year}{2022}\natexlab{}.
\newblock \showarticletitle{Minority salience and the overestimation of individuals from minority groups in perception and memory}.
\newblock \bibinfo{journal}{\emph{Proceedings of the National Academy of Sciences}} \bibinfo{volume}{119}, \bibinfo{number}{12} (\bibinfo{year}{2022}), \bibinfo{pages}{e2116884119}.
\newblock


\bibitem[Lee et~al\mbox{.}(2019)]%
        {lee2019homophily}
\bibfield{author}{\bibinfo{person}{Eun Lee}, \bibinfo{person}{Fariba Karimi}, \bibinfo{person}{Claudia Wagner}, \bibinfo{person}{Hang-Hyun Jo}, \bibinfo{person}{Markus Strohmaier}, {and} \bibinfo{person}{Mirta Galesic}.} \bibinfo{year}{2019}\natexlab{}.
\newblock \showarticletitle{Homophily and minority-group size explain perception biases in social networks}.
\newblock \bibinfo{journal}{\emph{Nature human behaviour}} \bibinfo{volume}{3}, \bibinfo{number}{10} (\bibinfo{year}{2019}), \bibinfo{pages}{1078--1087}.
\newblock


\bibitem[Lerman et~al\mbox{.}(2016)]%
        {lerman2016majority}
\bibfield{author}{\bibinfo{person}{Kristina Lerman}, \bibinfo{person}{Xiaoran Yan}, {and} \bibinfo{person}{Xin-Zeng Wu}.} \bibinfo{year}{2016}\natexlab{}.
\newblock \showarticletitle{The" majority illusion" in social networks}.
\newblock \bibinfo{journal}{\emph{PloS one}} \bibinfo{volume}{11}, \bibinfo{number}{2} (\bibinfo{year}{2016}), \bibinfo{pages}{e0147617}.
\newblock


\bibitem[Mansoury et~al\mbox{.}(2021)]%
        {mansoury2021graph}
\bibfield{author}{\bibinfo{person}{Masoud Mansoury}, \bibinfo{person}{Himan Abdollahpouri}, \bibinfo{person}{Mykola Pechenizkiy}, \bibinfo{person}{Bamshad Mobasher}, {and} \bibinfo{person}{Robin Burke}.} \bibinfo{year}{2021}\natexlab{}.
\newblock \showarticletitle{A graph-based approach for mitigating multi-sided exposure bias in recommender systems}.
\newblock \bibinfo{journal}{\emph{ACM Transactions on Information Systems (TOIS)}} \bibinfo{volume}{40}, \bibinfo{number}{2} (\bibinfo{year}{2021}), \bibinfo{pages}{1--31}.
\newblock


\bibitem[Marwick and Boyd(2011)]%
        {marwick2011tweet}
\bibfield{author}{\bibinfo{person}{Alice~E Marwick} {and} \bibinfo{person}{Danah Boyd}.} \bibinfo{year}{2011}\natexlab{}.
\newblock \showarticletitle{I tweet honestly, I tweet passionately: Twitter users, context collapse, and the imagined audience}.
\newblock \bibinfo{journal}{\emph{New media \& society}} \bibinfo{volume}{13}, \bibinfo{number}{1} (\bibinfo{year}{2011}), \bibinfo{pages}{114--133}.
\newblock


\bibitem[Mehrabi et~al\mbox{.}(2021)]%
        {mehrabi2021survey}
\bibfield{author}{\bibinfo{person}{Ninareh Mehrabi}, \bibinfo{person}{Fred Morstatter}, \bibinfo{person}{Nripsuta Saxena}, \bibinfo{person}{Kristina Lerman}, {and} \bibinfo{person}{Aram Galstyan}.} \bibinfo{year}{2021}\natexlab{}.
\newblock \showarticletitle{A survey on bias and fairness in machine learning}.
\newblock \bibinfo{journal}{\emph{ACM computing surveys (CSUR)}} \bibinfo{volume}{54}, \bibinfo{number}{6} (\bibinfo{year}{2021}), \bibinfo{pages}{1--35}.
\newblock


\bibitem[Santos et~al\mbox{.}(2021)]%
        {santos2021biased}
\bibfield{author}{\bibinfo{person}{Fernando~P Santos}, \bibinfo{person}{Simon~A Levin}, {and} \bibinfo{person}{V{\'\i}tor~V Vasconcelos}.} \bibinfo{year}{2021}\natexlab{}.
\newblock \showarticletitle{Biased perceptions explain collective action deadlocks and suggest new mechanisms to prompt cooperation}.
\newblock \bibinfo{journal}{\emph{Iscience}} \bibinfo{volume}{24}, \bibinfo{number}{4} (\bibinfo{year}{2021}).
\newblock


\bibitem[Sparkman et~al\mbox{.}(2022)]%
        {sparkman2022americans}
\bibfield{author}{\bibinfo{person}{Gregg Sparkman}, \bibinfo{person}{Nathan Geiger}, {and} \bibinfo{person}{Elke~U Weber}.} \bibinfo{year}{2022}\natexlab{}.
\newblock \showarticletitle{Americans experience a false social reality by underestimating popular climate policy support by nearly half}.
\newblock \bibinfo{journal}{\emph{Nature communications}} \bibinfo{volume}{13}, \bibinfo{number}{1} (\bibinfo{year}{2022}), \bibinfo{pages}{4779}.
\newblock


\bibitem[Wu et~al\mbox{.}(2022)]%
        {wu2022joint}
\bibfield{author}{\bibinfo{person}{Haolun Wu}, \bibinfo{person}{Bhaskar Mitra}, \bibinfo{person}{Chen Ma}, \bibinfo{person}{Fernando Diaz}, {and} \bibinfo{person}{Xue Liu}.} \bibinfo{year}{2022}\natexlab{}.
\newblock \showarticletitle{Joint multisided exposure fairness for recommendation}. In \bibinfo{booktitle}{\emph{Proceedings of the 45th International ACM SIGIR Conference on research and development in information retrieval}}. \bibinfo{pages}{703--714}.
\newblock


\end{thebibliography}


\end{document}